\pdfoutput=1
\documentclass[apjl]{emulateapj}

\usepackage{epstopdf}
\usepackage{lineno}
\usepackage{xspace}
\usepackage{graphicx}
\usepackage{amssymb}
\usepackage{amsmath}
\usepackage{natbib}

\newcommand{\fermi}{\textsl{Fermi}\xspace}

\shorttitle{Detection of QPOs in PKS 0219-164}
\shortauthors{Bhatta et al.}

\begin{document}

\title{Radio and $\gamma$-ray variability in the BL Lac PKS 0219 -164:  Detection of Quasi-Periodic Oscillations in the Radio light curve.}

\author{
Gopal~Bhatta\altaffilmark{1}
}

\altaffiltext{1}{Astronomical Observatory of the Jagiellonian University, ul. Orla 171, 30-244 Krak\'ow, Poland}
\altaffiltext{2}{Mt. Suhora Observatory, Pedagogical University, ul. Podchorazych 2, 30-084 Krak\'ow, Poland}

\email{email: {\tt gopalbhatta716@gmail.com}}

\begin{abstract}
 In this work, we explore the long-term variability properties of the blazar PKS 0219-164 in the radio and the $\gamma$-ray regime, utilizing the OVRO 15 GHz and the \fermi/LAT observations from the period 2008--2017.  We found that $\gamma$-ray emission is more variable than the radio emission implying that $\gamma$ ray emission possibly originated in more compact regions while the radio emission represented continuum emission from the large scale jets. Also, in $\gamma$-ray the source exhibited spectral variability characterized by the \emph{ softer-when-brighter} trend, a less frequently observed feature in the high energy emission by BL Lacs. In radio, using Lomb-Scargle periodogram and weighted wavelet z-transform, we detected a strong signal of quasi-periodic oscillation (QPO) with a periodicity of 270 $\pm$ 26 days with possible harmonics of 550 $\pm$ 42 and  1150 $\pm$ 157 days periods. At a time when detections of QPOs in blazars are still under debate, the observed QPO with high statistical significance ( $\sim$ 97\% -- 99\% global significance over underlying red-noise processes) and persistent over nearly 10 oscillations could make one of the strongest cases for the detection of QPOs in blazar light curves.  We discuss various blazar  models that might lead to the $\gamma$-ray and radio variability, QPO, and the achromatic behavior seen in the high energy emission from the source.
\end{abstract}


\section{Introduction \label{sec:intro}}

Active galactic nuclei (AGN) are the brightest sources in the universe widely accepted to be powered by accretion on to supermassive black holes. Based on their radio continuum, they can be broadly classified as radio-loud and radio-quiet objects: the sources with the ratio of nuclear radio emission at 5 GHz to the optical emission at (4400 $\AA$) greater and less than 10, respectively \citep{Kellermann1989}. Most of the radio-loud AGNs display relativistic jets originating close to the central region and extending up to Mpc scale \citep{Fanaroff1974}. Blazars are the subclass of radio-loud AGNs with their jets aligned close to the line of sight such that the relativistic effects become pronounced leading to the Doppler boosted emission that is highly variable over the entire electromagnetic spectrum \citep{up95}. The broadband spectral energy distribution (SED) of blazars can often be recognized by the double-peaked featured in the $\nu$--$\nu F_{\nu}$ representation. The lower peak, usually found between the radio and the X-ray, is attributed to the synchrotron emission by the energetic particles -- electrons in leptonic models  e.g., \citet{Maraschi1992} and \citet{Bloom1996}, and protons in hadronic models e.g., \citet{Mannheim1992}, \citet{Aharonian2000} and \citet{Mucke2003} -- encircling the jet magnetic field; the other peak mostly lying between UV to $\gamma$-ray is resulted due to the inverse-Compton scattering of the soft seed photons by the energetic particles accelerated by various mechanisms. In such case, the seed photons might originate at the various components of an AGN e.g.  accretion disk \citep{Dermer1993}, broad-line region (BLR; \citet{Sikora1994} and dusty torus \citep{Blazejowski2000}. Some of the widely discussed particle acceleration  scenarios producing highly energetic particles include impulsive electron injections near the base of a jet \citep[e.g.][]{Dermer1997}, internal shocks propagating along the jets \citep[e.g.][]{Marscher1985,Kirk98,Sokolov2004}, stochastic particle acceleration in shear boundary layer of the relativistic jets \citep[see][]{Ostrowski2002} and particle acceleration due to the turbulence in the relativistic jets \citep[see][and the references therein]{Marscher2014}.

Blazars further consist of two kinds of sources: flat-spectrum radio quasars (FSRQ) and  BL Lacertae (BL Lac) objects. FSQRs are more powerful, show emission lines over the continuum and have the synchrotron peak in the lower part of the spectrum; whereas BL Lac objects are less powerful, show weak or no emission lines and have synchrotron peak in the higher part of the spectrum. BL Lacs represent an extreme class of sources with maximum synchrotron and inverse-Compton energies (hard X-rays to TeV emission); however, in comparison to the more luminous FSRQs, they accrete at relatively low rates, and  do not possess strong circum-nuclear photon fields.

\begin{figure*}[!t]
\begin{center}
\includegraphics[width=0.65\textwidth,angle=-90]{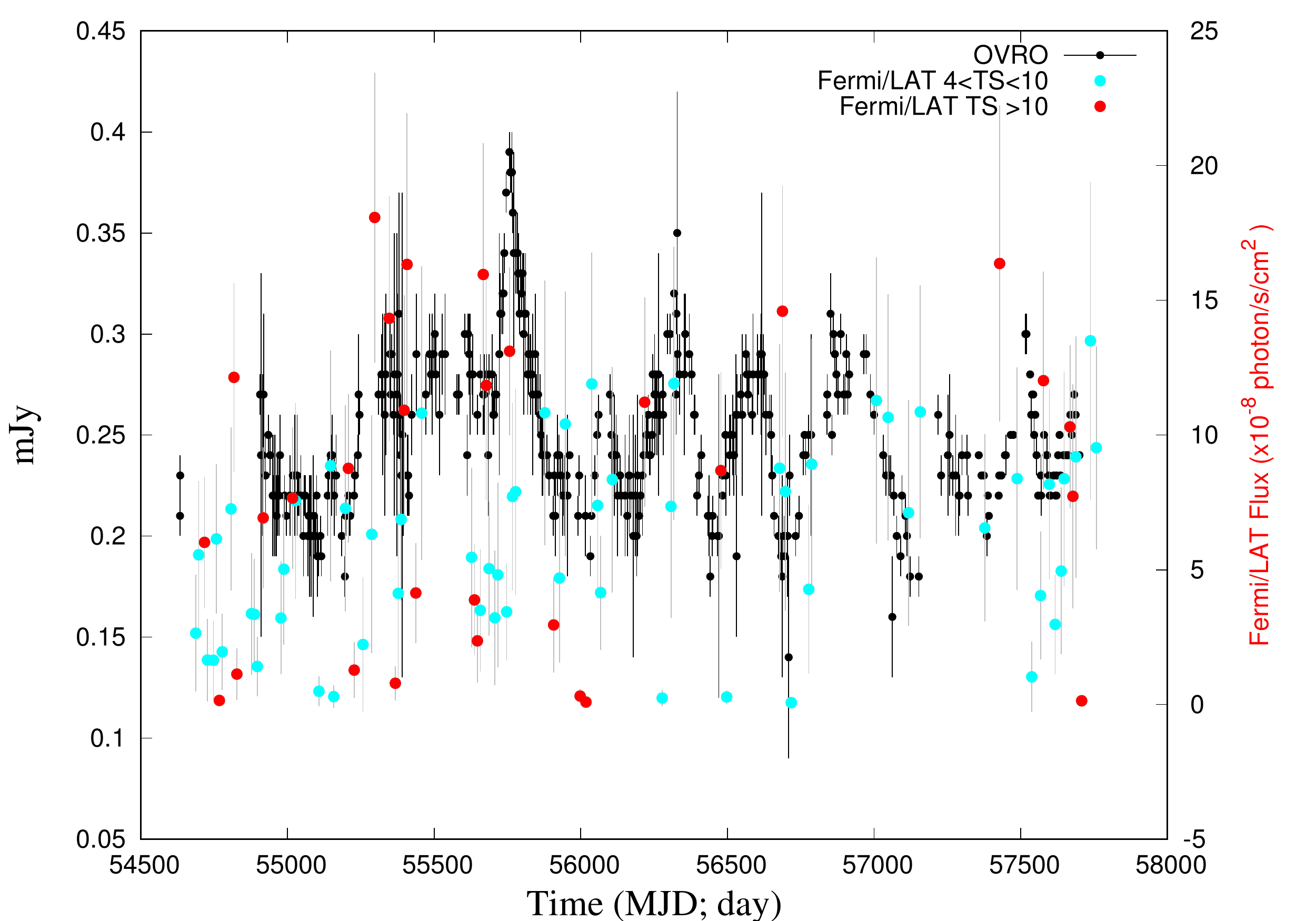}
\caption{The 15\ GHz OVRO  observations of the source PKS 0219-164 are shown by the black symbols. The \fermi/LAT  (0.1--300 GeV) observations are shown by red (TS $>$ 10) and  cyan ($4<$TS$<10$) symbols. For clarity, the measurement errors associated with the red and cyan symbols are shown in gray color. \label{Fig:1}}
\end{center}
\end{figure*}

Blazar continuum emission is characterized by broadband emission  which is variable, both flux and polarization, on diverse timescales. The variability timescales can be broadly classified as long-term, short-term and intraday/night variability. Longterm variability, typically in the timescale of a few years, might arise due to variable accretion rates; short-term variability, which are usually identified with flaring episodes lasting a few weeks to a few months, could be the result of the passage of the shock waves propagating down the blazar jets; and the intraday variability might  be resulted due to the turbulence in the innermost region of the jets  \citep[e.g.][]{Cawthorne2006,Lister2005,Hughes98,Marscher1996}

 In general, the variability shown by AGNs appears predominantly aperiodic in nature. However, in recent times  reports of QPOs in the multi-frequency light curves of AGNs have begun to accumulate.  Generally, signatures of (quasi-)periodic  oscillations can be revealed as the large peaks in the periodograms of the source light curves. QPOs have been claimed to have been found in various frequencies, on diverse characteristic timescales, and in various classes of AGNs; some of them are listed below. 

\begin{itemize}
  \item   QPOs in the radio light curves of AGNs were reported to show periodicities in the timescales of  a few years,  \citep[e.g.][]{Wang14,An13,Hovatta08,Xie08,Liu06,Raiteri01}.
  \item    QPOs were detected in the optical light curves of blazars  with periodicities in the time scales of  a few years   \citep[e.g.][]{Sandrinelli16,bhatta16c},
   \item  QPOs in  X-ray light curves of AGNs were reported with periodicities in the time scales a few hours \citep[e.g.][]{Lachowicz09,Gierlinski08}
    \item  QPOs were found in the $\gamma$-ray light curves of  blazars with periodicities in the time scales a few years  \citep[e.g.][]{Zhang2017,Ackermann2015,Sandrinelli14}   
     \item QPOs were reported in the light curves of radio-quiet quasars \citep[e.g.][]{Zheng16,Graham15a,Graham15b,King13}
     \item QPO was detected in high-redshift (z = 2) radio-loud quasar  \citep[][]{Liu15}
\end{itemize}

\noindent From above it is clear that QPOs do not seem to prefer any particular system or timescale; rather they could be found in a wide range of AGN classes.

   PKS\ 0219-164 (R.A.=$\rm 02^{h}22^{m}00.7^{s}$, Dec.= $-16\degr15\arcmin17\arcsec$, and $z=0.7$) is   a BL Lac reported to have been observed over a broad range of the electromagnetic spectrum: e. g., optical \citep{Ballard1990}, infra-red \citep{Impey1988}, infra-red and optical polarization \citep{Mead1990} and radio 5 GHz observations \citep{Kharb2004}.  \citet{Condon1977} first estimated the accurate position of the source at radio frequency (2700 MHz) along with the position of its optical counterpart, and later the source was classified as quasi-stellar object (QSO) by \citet{Hewitt1993}.  In their observations, \citet{Meisenheimer1984} found the optical emission to be strongly polarized, up to as high as $\sim$ 19 \%, and highly variable. Similarly, the flux was also displaying a dramatic variability, changing its brightness by $\sim$ 3 magnitudes within a period of a week. An upper limit of 0.23 \% circular polarization was estimated during  a 5 GHz survey of the parsec-scale polarization properties of 40 active galactic nuclei (AGNs) made with the Very Long Baseline Array (VLBA) \citep{Homan2001}. An upper limit of the TeV range flux, 1.8 $\times\ 10^{-12}$ photons/cm$^{2}$/s, was measured by High-Energy-Gamma-Ray Astronomy (HEGRA) Cherenkov telescopes \citep{Aharonian2004}.  In the cross-correlation of the Fermi 11-month survey (1FGL) catalog with the 20-GHz Australia Telescope Compact Array (AT20G) radio survey catalog,  the radio and $\gamma$-core of the source, separated by 4.09 arcmin,  were found to be highly correlated with a probability of 0.95 \citep{Ghirlanda2010}. Also included in a very sensitive 21 cm survey \citep{Lockman2002}, the source is being  monitored regularly by 15\,GHz 40-m telescope at the Owens Valley Radio Observatory (OVRO; \citealt{Richards2011}). The source is listed as `3FGL J0222.1--1616'  in the \fermi/LAT 3rd catalog \citep{Ackermann2015}

 In this paper, we present our analysis and results of the radio and $\gamma$-ray variability study of the BL Lac source PKS 0219-164 using the observations from OVRO and \fermi/LAT  spanning $\sim 9$ years (section 2). In particular, we analyzed the long-term OVRO light curve using the Lomb-Scargle periodogram (LSP) and the weighted wavelet z-transform (WWZ) methods.  We report the detection of statistically significant  quasi-periodic oscillations in the flux with a characteristic timescale of 270 $\pm$ 26 days along  with possible low-frequency harmonics at the periods of 550 $\pm$ 42 and  1150 $\pm$ 157 days.  Using Monte Carlo (MC) simulations, the global significance of the detection over underlying red-noise processes was found to be  97.5\% --99.6\%. In addition, we observed an interesting spectral behavior in the  \fermi/LAT observations which displayed {\emph softer-when-brighter trend} (section 3). Finally we present our  discussion and conclusions in section 4.

\begin{figure}[]
\begin{center}
\includegraphics[width=\columnwidth]{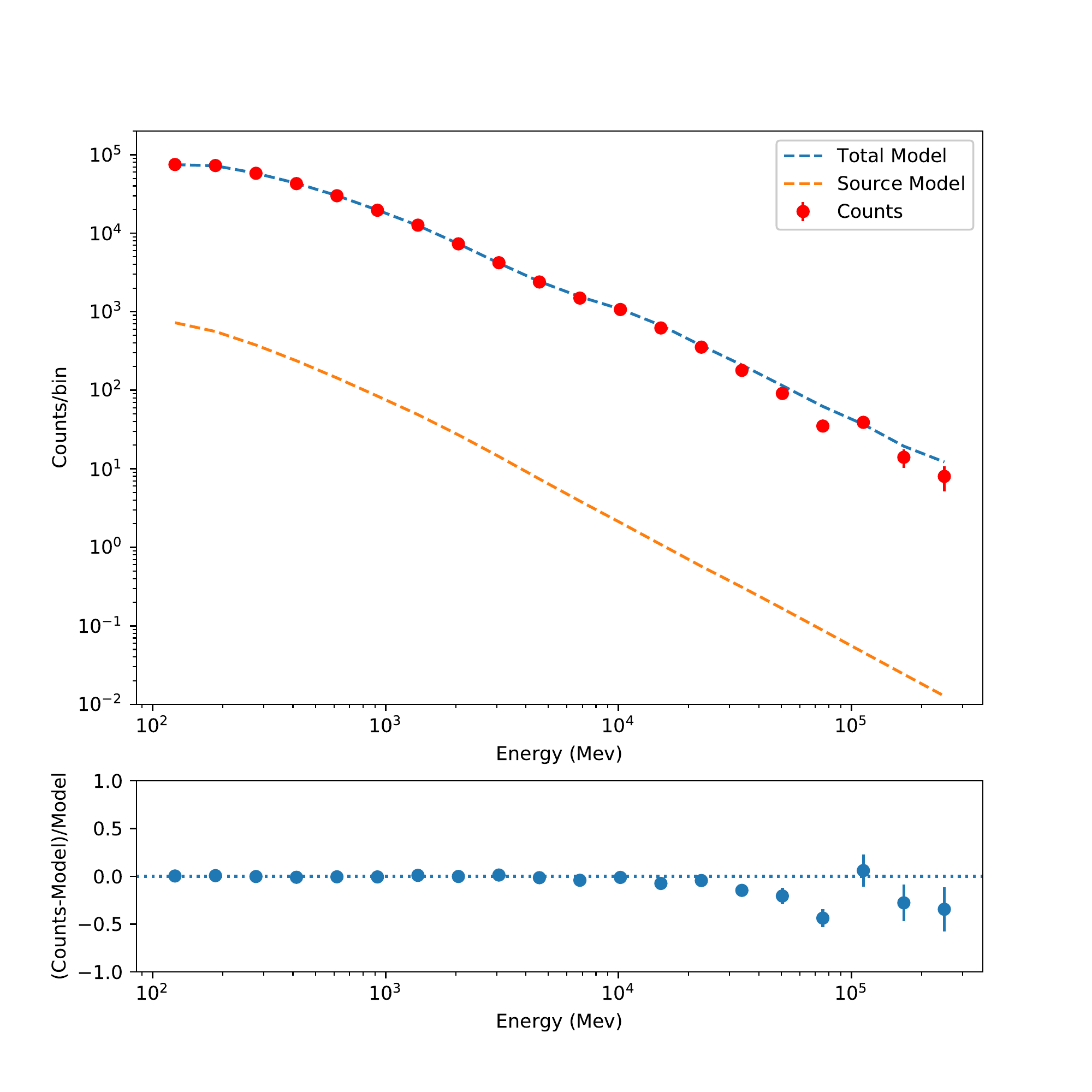}
\caption{Upper panel: Number of total source counts per energy bin (red symbols). The blue dashed line represents the sum of all the models and the orange line is the source model. Error bars on the red symbols represent square root of the observed number of counts in that bin. Lower panel: Residuals are computed between the total model and the total observed counts. \label{Fig2}}
\end{center}
\end{figure}

\begin{figure}[]
\begin{center}
\includegraphics[width=\columnwidth]{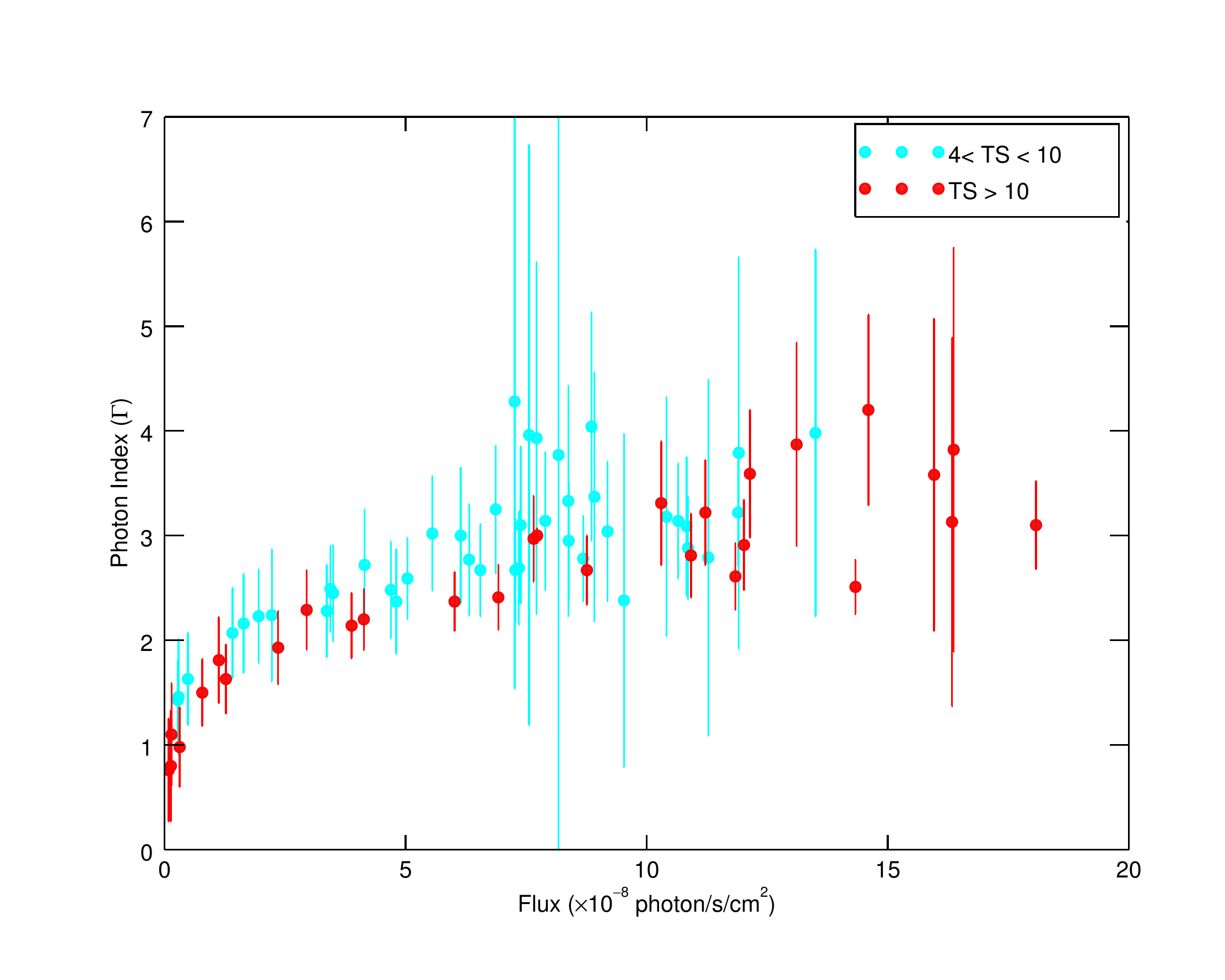}
\caption{The \fermi/LAT photon indexes are plotted against the simultaneous fluxes of the source PKS 0219-164. The red symbols represent the observations with test statistics (TS) $>$ 10 and the cyan ones with 4$<$TS$<10$ \label{Fig3}}
\end{center}
\end{figure}

\section{Observations and Data Reduction \label{sec:obs}}

\subsection{Radio Observations}

The 15\,GHz radio observations of the source PKS 0219 -164 were obtained from the AGN monitoring program at the Owens Valley Radio Observatory using the 40-meter telescope. The radio  light curve from the observation epoch 54633 -- 57700 MJD is presented in Figure \ref{Fig:1}.

\subsection{\fermi/LAT Observations }
 The Large Area Telescope (LAT) of the Fermi Gamma-ray Space Telescope (\fermi/LAT) is a all-sky monitoring instrument operating between 20 MeV to 300 GeV energy range \citep{Atwood2009}. With its large effective area ($> 8000\ cm^{-2}$), wide field of view ($>$ 2 sr) and high energy resolution ($ < 3.5^{o}$  around 100 MeV and $< 0.15^{o}$  above 10 GeV), the telescope is one of the most useful instruments in the study of the universe in high energy. \fermi/LAT observations of the source PKS 0219 -164 (or 3FGL J0222.1-1616) were  processed using the Fermi Science Tools along with the \fermi/LAT catalog, Galactic diffuse emission model and isotropic model for point sources; and the standard procedures of the unbinned likelihood  analysis were followed.\footnote{https://fermi.gsfc.nasa.gov/ssc/data/analysis/scitools/} The observations from the period 2008-08-04 -- 2017-01-23 (equivalently 54682 -- 57776 MJD) were considered for the analysis. First, selections of the events were made, using the \fermi tool \emph{ gtselect}, selecting only the events in a circular region of interest (ROI) of $10^{\circ}$ radius centered around the source, and then limiting the zenith angle $ <$ 90$^{\circ}$ to minimize the contamination of $\gamma$-rays from the Earth limb. Subsequently, tool \emph{gtmktime} was used to select the good time intervals (GTI) to ensure that the satellite was operating in the standard science mode resulting in good  quality of the observations. After making an exposure map using \emph{gtexpmap} and \emph{gtltcube}, a source model file containing model parameters and the source positions for all the sources within the ROI was created  using the Python application make3FGLxml.py. Then the diffuse source response  was calculated using the Galactic and extragalactic models of the diffuse $\gamma$-ray emission such as \emph{ gll iem v06.fit} and \emph{ iso\_P8R2 SOURCE V6 v06.txt}. Finally, likelihood ratio test \citep{Mattox1996} was performed (using \emph{ gtlike}) to estimate the significance of the $\gamma$-ray events from the source.  With the set of the parameters given in the input source models, the task attempts to  maximize the probability density of the data (given the current model) by fitting all the sources within the ROI.  The maximum probability density of the data under two different models is then compared in the likelihood ratio test statistic given by $TS = 2×(logL_{1} - logL_{0})$, where L$_{1}$ and L$_{0}$ represent   the likelihood of the optimized parameters given the data, under the null and alternative model, respectively.  Then the significance of a source detection is expressed by $\sim \sqrt{TS} \sigma$ \citep{Abdo2010}. 
 
We first analyzed the entire data to estimate the average flux and spectrum using a power-law model for the input source model.  The resulting TS value, average photon index were  296.8 and $2.59 \pm 0.06$, respectively, and the  average photon flux was found to be $1.58159^{-08} \pm 1.5453^{-09}$ photons/cm$^{2}$/s. These values are in close agreement with the values from 3FGL catalog. The likelihood analysis resulted in the count distribution for all the sources in the ROI which, along with the total model (blue dashed curve) and the source model (orange dashed curve), is shown by the red symbols in Figure  \ref{Fig2} . 

We then generated light curves using three different kinds of time bins i. e., 10, 15 and 30 d  bins. But 10 d bin resulted maximum number (although only 31 for the total observation period) of observations with TS value $>$ 10. Although, for a robust conclusion only the observations with TS value $>$ 10 are considered, for a tentative trend the observations with $4<$TS$<10$  will also be considered in the following analyses. 

\begin{figure}[]
\begin{center}
\includegraphics[width=\columnwidth]{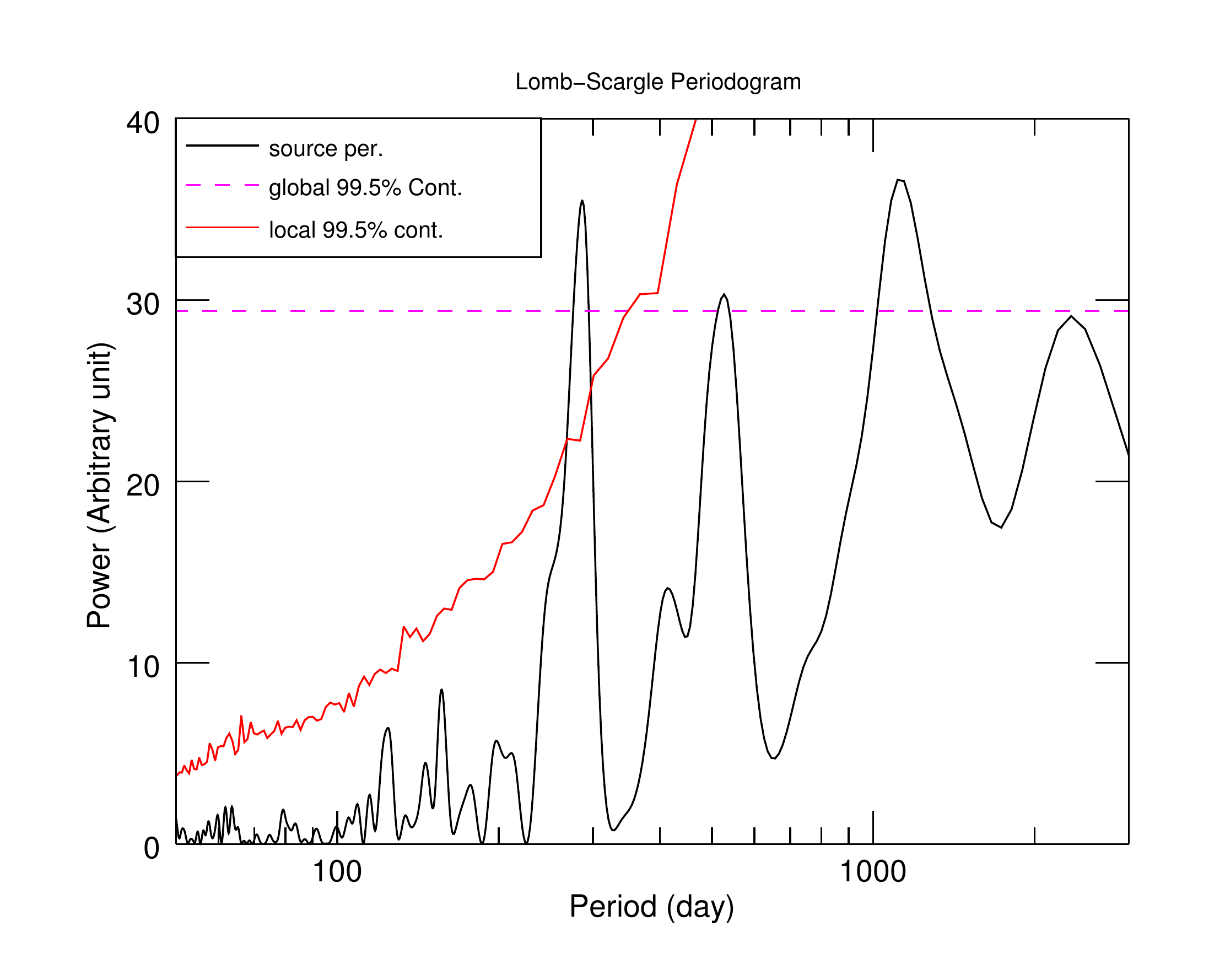}\\
\caption{The Lomb-Scargle periodogram of the OVRO light curve of the source PKS 0219-164 is represented by the black curve. The red curve  and the magenta line represent the 99.5\% local and global significance contours, respectively,  from the MC simulations. \label{Fig4}}
\end{center}
\end{figure}

\section{Analysis and Results \label{sec:dis}}

\subsection{Radio and $\gamma$-ray  Variability}
Figure \ref{Fig:1} shows the longterm \fermi/LAT and OVRO light curves of the blazar PKS 0219 -164. The red symbols in the $\gamma$-ray light curve represent the \fermi/LAT detections with TS $>$ 10, and therefore the results from the analysis of such observations will be considered more robust. However, as there are considerably fewer number of observations with TS $>$ 10, the observations with 4$<$ TS $<$ 10 are also presented.  From the figure it can be seen that both radio and $\gamma$-ray emission display considerable variability during the observation period. As a quantitative measurement of the variability,  fractional variability \citep{vau03,Aleksic2015} for the radio and the $\gamma$ ray light curves are estimated to be  $0.14 \pm 0.01$  $0.51\pm0.05$, respectively. The observed more rapid variability and the higher fractional variability in $\gamma$ ray emission suggests that the source is more variable in $\gamma$ ray than in the radio emission. The observation is consistent with the fact that the SEDs of BL Lacs peak in the higher part of the spectrum (e. g., in X-ray and TeV range).

\begin{figure*}[!t]
\begin{center}
\includegraphics[width=\textwidth]{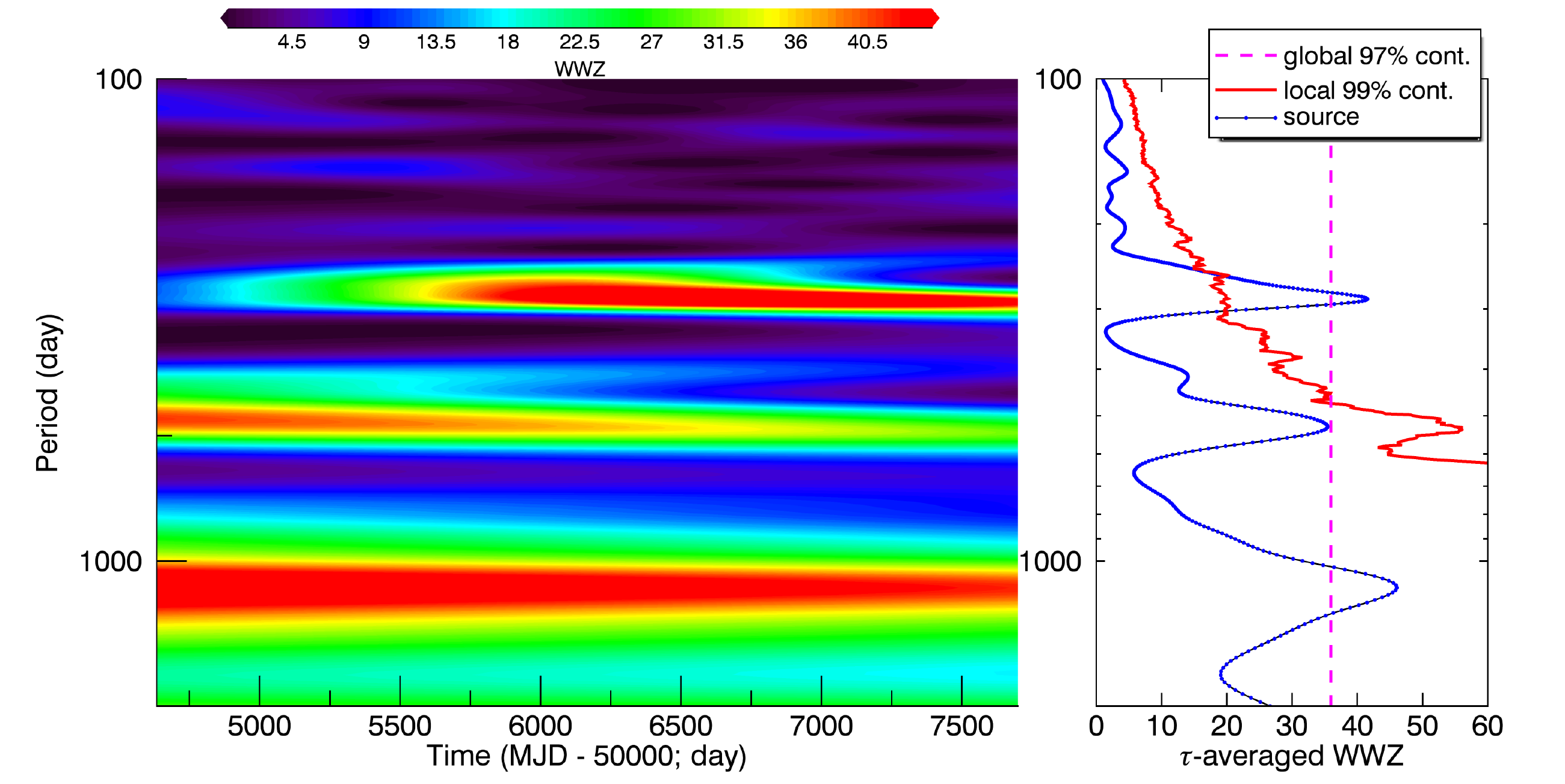}
\caption{Weighted wavelet z-transform  of  OVRO light curve for the blazar PKS 0219-164. The left panel shows the distribution of color-scaled WWZ power in the time-period plane, and the right panel shows the time averaged power (blue curve) as a function of period. The red curve and the magenta line show the 99\% local significance and 97\% global significance contours, respectively,  from the MC simulations. \label{Fig5}}
\end{center}
\end{figure*}

\subsection{Spectral Variability: Softer-when-brighter Trend}

Study of correlation between  simultaneous variability in flux and spectrum offers an important insight into the blazar physics, in particular the particle acceleration and the energy dissipation mechanisms. In this context, we find some observational properties that characterize the correlation between the flux and the spectral slope of AGNs. In optical, the anti-correlation between them has been observed in the form of  the \emph{bluer-when-brighter} trend in some BL Lacs  \citep[see][]{Wierzcholska15,Ikejiri11,Sasada08,Wu2007,Villata2004,Villata2002,Raiteri01}; whereas in some FSRQs an correlation between the flux and the spectral index has been reported in the form of  \emph{redder-when-brighter} trend  \citep[e.g.][]{Bonning2012,Raiteri2008,Villata2006}.  Similarly, in hard X-ray harder-when-brighter was observed in Tev blazars \citep{Pandey2017}. 

 To explore similar correlation in the source, $\gamma$-ray photon indexes were plotted against the corresponding simultaneous fluxes as shown in Figure \ref{Fig3}. A large value of Spearman's correlation coefficient ($\sim$ 0.90)   implies a strong correlation between the spectral index and the source intensity suggesting a presence of {\emph softer-when-brighter} trend.  In the figure, the trend is also traced by the less significant observations (with 4$<$TS $<$ 10) as shown by the cyan symbols. In $\gamma$-ray, we find more reports of harder-when-brighter cases \citep[][and the references therein]{Brown2011} than that of the softer-when-brighter ones \citep[e.g.][]{Foschini2010}. Therefore the observed softer-when-brighter  feature in the BL Lac could be potentially important in the understanding of the spectral behavior of blazars in the high energy regime.

\subsection{Periodicity Search: Lomb-Scargle Periodogram and  Weighted Wavelet Z-transform}

The search for the presence of QPOs in the OVRO light curve was conducted using Lomb-Scargle periodogram (LSP) \citep{Lomb76,Scargle82} and weighted wavelet z-transform (WWZ) which are frequently used methods in the astronomical time series analysis. The LSP method, although a form of the traditional discrete Fourier transform (DFT), has advantages over DFT  that it reduces the effect of irregular sampling by the least-square fitting of the sinusoidal waves to the data.   Consequently, such fitting increases the significance of the observed power spectral features (or peaks) which can potentially represent QPO signals present in the light curves. The  LSP powers of the  OVRO light curve of the source for the observation period considered are presented in Figure~\ref{Fig4}. The figure shows three distinct peaks  around the periods $270\pm26$, $550\pm42$ and $1150\pm157$ days. The uncertainties in the period are represented by the half-width at the half-maximum (HWHM) of the peaks. 

Although the LSP method is suitable for the periodicity analysis of the light curves with irregular sampling, the method tries to fit the waves throughout the entire dataset. In other words,  it does not consider the fact that in the real astronomical systems the periodic oscillations  may evolve over time; i.e., frequency and amplitude of the oscillations may change over time. In such situations, the wavelet transform method becomes a more suitable tool to investigate the presence of the QPOs which develop and decay with time. The method, like the LSP, also attempts to fit sinusoidal waves to the data; however the waves can be localized in both time and frequency domains to account for the transient nature of QPOs \citep{Bravo2014,Torrence98}. The method  has been widely used in the time series analysis of blazar light curves\citep[e.g.,][]{bhatta16c,Mohan15,bhatta13,Hovatta08}.  To look for the possible QPOs in the radio light curve of the blazar PKS 0219-164, we employed  weighted wavelet z-transform (WWZ), a method described in \citet{Foster96}.  Using the WWZ software,\footnote{\texttt{https://www.aavso.org/software-directory}} we estimated the WWZ power of the light curve as a function of time and period. The color-scaled WWZ powers of the source light curve in the time-period plane  are presented in Figure~\ref{Fig5} which reveals large WWZ power centered around the periods 270, 550 and 1150 days implying the presence of the possible QPOs with the corresponding periods. The QPO with the period of 270 days  appears to gradually develop from the start of the observation epoch and grows stronger towards the end of the observations; the QPO corresponding to the period of 550 days, most likely the first harmonics gradually decays before the end of observations. On the other hand, the signal corresponding to the period of 1150 days seems persistent throughout the observation period.  The right panel of Figure~\ref{Fig5} shows the time-averaged WWZ power at a given period. Once again, in the panel, we can see three distinct peaks centered at the periods of $270\pm21$, $550\pm52$ and $1150\pm182$ days. As in the LSP, HWHMs about the central peaks provide a measure of the uncertainties in the observed periods. 

\subsection{Significance Estimation: MC Simulations}

Due to the noisy nature of the periodogram, large spurious peaks can be mistaken as the periodic signals. Therefore it is important that we carefully analyze the effect of the uneven sampling on the noisy behavior of the periodogram. In addition, blazar variability, in general, exhibits red-noise like behavior such that the periodograms can be well represented by a power spectral density (PSD) of the form $P(\nu) \propto \nu ^{-\beta}$; where ${\nu}$ and  ${\beta}$ represent temporal frequency and spectral slope \citep[see][]{Max-Moerbeck2014}. This means that the light curves showing such power-law like PSD  oscillate with large amplitudes on longer timescales, and hence there is always a chance that these oscillations can also be mistaken as  QPO features \citep[see][]{Vaughan2016,Press78}. Therefore, for a rigorous estimation of the significance of the QPO features revealed by the LSP and WWZ methods,  both of these issues should be taken into consideration.  In this work, these issues are addressed by studying the periodograms and WWZ powers of a large number of simulated light curves that have the same sampling as the of the source light curve. For the purpose, the Monte Carlo (MC) simulations of the light curves were performed by randomizing the amplitude as well as the phase of the Fourier components \citep[for details see][]{TK95}.

Now, to estimate the power spectral shape of the underlying colored noise in the light curve of the source, we followed the power response method  \citep[PSRESP;][]{Uttley02} - one of the widely used methods to characterize AGN power spectral density \citep[e.g.,][]{Chatterjee08,Edelson2014,bhatta16b,bhatta16c}.To evaluate the best-fit PSD model, the method calculates the probability that the source  periodogram best represents a given model PSD.  Various model PSDs with varying parameters (e.g., spectral slope in our case) are fitted the binned source periodogram to obtain the highest probability. The resulting best-fit model PSD is then used to simulate the light curves for further analysis.  Using the best-fit PSD model, 10000 light curves were simulated and subsequently re-sampled according to the sampling of the source light curve.  The spectral distribution of the simulated light curves was analyzed to evaluate the \textbf{local} significance of the QPO features seen in the observed LSP.  In particular,  the local 99.5\% significance (99.5 percentile) contour, shown by the red curve in Figure~\ref{Fig4}, were determined \citep[for further details see][]{bhatta16b,bhatta16c}. As seen from the figure, the local significance of the observed QPO feature at the period of 270 d turns out to be greater than 99.5\%. 

The above method of estimating the significance of a LSP peak provides \emph{local} estimates as it represents the significance only at a particular period.  Since we do not have \textsl{a priori} knowledge of where the significant peak might occur, a more robust  measure of the significance can be \emph{global significance} which is associated with the fraction of the simulated LSP powers  at any period below the observed power at the period of our interest  \cite[see][]{Bell2011}. We evaluated the global significance of the LSP peak at the period 270 to be 99.6 \%. These large values of both local and global significances indicate that the spectral peak at 270 d represents a real and physical periodic signal as opposed to the oscillations due to the underlying red-noise processes.  However for the spectral features at the periods of 550 and 1110 days, the significances look much smaller.

We took a similar approach to evaluate the significance of the observed WWZ features at 270 d, 550 d and 1110 d periods. We first simulated 10000 light curves from the best-fitting PSD model and re-sampled them according to the source light curve. Subsequently,  wavelet analysis was performed to calculate time averaged WWZ  power for a given period.  In particular, the time ($\tau$) averaged WWZ power of the source was compared against the  99\% local significance and 97\% global significance contours derived from the distribution of the time averaged WWZ power of the simulated light curves. In the right panel of  Figure \ref{Fig5}, the 99\% local significance and 97\% global significance contours are shown by the red curve and the magenta line, respectively. As before, with the estimated global significance of  97.5\%, the 270 d QPO feature turns out to be highly significant. However, as in LSP, the significances  for the possible harmonics at  550 d and 1110 d periods are observed to be much lower.

\section{Discussion and Conclusion \label{discussion}}

\subsection{Radio and $\gamma$ ray Variability}

Multi-frequency variability in blazars over various timescales could be resulted from a combination of modulations in the emission originating at various geometrical components including accretion disk, jets, dusty torus, BLR etc. In particular, the origin of longterm  radio variability in blazars could primarily be  associated with the synchrotron emission from the large scale jets. In blazars, the large scale radio jets are clearly resolved in the radio images; some of the radio knots appear to be moving with superluminal motion with apparent velocity up to $\sim$ 37c \citep{Jorstad01}. This indicates that the variability in the radio continuum most likely arises due to the modulations in the radio emission from the jets. 

As for the origin of the $\gamma$-ray emission, there appears to be an uncertainty about its exact location relative to the central engine. Given the high energy activity and rapid variability in the timescales of a few minutes (e.g., \citealt{Ackermann2016} for 3C\,279 and  \citealt{Aharonian2007} for PKS\,2155-304), $\gamma$-ray emission could  originate at compact regions close to the central black hole. However  in such scenario the bulk Doppler factor has to be very large (typically $\delta > 60$)  to limit the $\gamma \gamma$ opacity in the rest frame below the threshold value above which the pair production takes over depleting $\gamma$ ray photons drastically. On the other hand, the emission could also originate near the millimeter-wave core, a few kiloparsecs from the central region \citep[in case of blazar OJ 287 see][]{Agudo2011}. But there seems to be an apparent lack of intense photon field required for the inverse-Compton process to boost the soft photons up to the $\gamma$-ray regime \citep[for further discussion see][]{Dermer2015}.

The modulation in the flux of blazars could arise due to the propagation of the relativistic shocks along the blazar jets viewed at small angles  \citep[e.g.][]{Marscher1985,Spada2001,Joshi2011}. In addition, the non-thermal emission variability can arise due to various instabilities in the jet e.g. turbulence behind the shocks. \citep[e.g.][]{bhatta13,Marscher2014}. 
Similarly, sometimes the disk-based hotspots or instabilities \citep[e.g.][]{Wagner95,up95,Chakrabarti 93} can propagate into the jet to modulate the physical parameters of the jet such as its velocity, density and magnetic field, which in turn can also result flux variability  \citep[e.g.][]{Wiita06}.

 Alternatively, such variations can be of extrinsic origin e.g., geometrical effects involving swing in the emission region about the line of sight.   For instance, even a slight change in viewing angle and/or bulk Lorentz factor leading to variation in Doppler factor can produce a large variations in observed Doppler boosted flux  by the relations $F_{\nu}=\delta^{3+\alpha}{F}'_{{\nu}'}$ and $\delta=1/\Gamma \left ( 1-\beta cos\theta  \right )$, where the bulk Lorentz factor, velocity, spectral index and the viewing angle are represented by $\Gamma$, $\beta c$, $\alpha$  and $\theta$, respectively;  and the primed quantities in the relation represent the quantities in the co-moving frame  \citep[][]{Blandford78,Gopal-Krishna03}. To illustrate the argument, a slight swing ($\theta$ $= 0.6^{o}$) in the viewing angle  made by  an emission region of moderate  $\Gamma =10$ with a spectral index of 1.6 ( spectral index for PKS 0219-164 equivalent to the \fermi/LAT photon index of 2.6) can produce a change of $\sim$ 50\% in flux, which is the order of the variability we see in the source in $\gamma$-ray band \citep[for detailed discussion refer to] []{Ghisellini97}.

\subsection{Achromatic Behavior}

Any correlation between  flux and spectral variability in a source implies a close connection between the observed flux enhancement and renewed particle injection in situ \cite[e.g.][]{Mastichiadis2002,Kirk98}.   However, in blazars the nature of the correlation between the flux and the spectral index, so far, is somewhat uncertain.  In the optical band, generally,  BL Lacs  have found to exhibit  bluer-when-brighter tendency, whereas FSRQs often show  redder-when-brighter trend. However,  several sources have been reported to exhibit both kinds of spectral behaviors depending on their flux state \citep[for recent study see][]{Acosta2017}. In $\gamma$-ray regime also  blazars  were found to behave in the similar fashion i.e., in some cases the spectrum hardens with the source intensity and in other cases  the spectrum softens with the flux enhancements. In addition, some sources show both correlation and anti-correlation between the  spectral index and the flux state  \citep[see][]{Nandikotkur2007}. In the particle acceleration scenario where the distribution of the injected particle is of the single power-law form, such multi-frequency achromatic behavior are hard to explain considering only purely geometrical or beaming effects of the emission regions.  But on the other hand, the correlation between spectral index and flux (i. e. softer-when-brighter similar to redder-when brighter in optical)  observed in the source PKS 0219-164  could be explained assuming an underlying steady electron energy spectrum of a curved (particularly convex) shape. In such case, if the dominant population of energetic particles participating in inverse-Compton process are near the lower energy of the spectrum, the resulting $\gamma$ emission can exhibit softer-when-brighter trend. Alternatively,  local enhancements of the magnetic field in the jets could  lead to synchrotron emission with an excess of hard photons, which further could be up-scattered by lower energy electrons to $\gamma$-ray  regime where they exhibit softer-when-brighter trend. However, for a robust and clear picture of the achromatic behavior to emerge out, further systematic study involving simultaneous multi-wavelength observations of blazar will be required.

\subsection{Quasi-periodic Variability}

Study of quasi-periodic oscillations sheds light into various aspects of AGN research such as strong gravity environment and disk-jet connection. In principle, (quasi-) periodic oscillations in AGNs can originate in various scenarios. The simplest among them is the emission region moving in the Keplerian orbits around the central black hole.  Other possible cases might include  periodicity of a binary black hole system \citep[e.g.,][]{Lehto96,Graham15a}, jet precession due to nearby massive object or warped accretion disks \citep[][]{Graham15b,Sandrinelli16}.   Similarly, jet modulation by various instabilities evolving near the innermost regions of accretion disks also could result in quasi-periodic modulation of the observed jet emission \citep[see][]{Liu06,Wang14}. In addition, globally perturbed  torus or a thick disk of finite radial extent could also give rise to QPOs \citep[e.g.][]{Rezzolla2003,Zanotti2003}.

 In the particular case of the blazar PKS 0219 -164, the detected QPO  with a periodicity of $\sim 270$\,days - along with the accompanying $\sim 510$\,d  and  $\sim 1150$\,d harmonics - could be explained in the context of highly magnetized jets and rotating magnetic field as suggested by \citet{Meisenheimer1984}. In such case, relativistic motion of the emission regions along the helical path of the magnetized jets could make a more plausible explanation \citep[e.g.,][]{Camenzind92, Mohan15}.  Similarly, in the magnetic flux paradigm for the jet launching in AGNs \citep[see][]{Sikora13}, the magnetic flux accumulation can lead to the formation of so-called \emph{magnetically choked accretion flow} (MCAF). In that case, the Rayleigh-Taylor and Kelvin-Helmholtz instabilities can cause QPO oscillations at the interface of the disk and the magnetosphere due to sudden change in the density and the magnetic flux \citep{Li2004,Fu2012}. The periodicities for these QPOs could range from a few days to a few months depending upon the black hole spin parameter. Furthermore, QPOs are also found to develop in recent magneto-hydrodynamical (MHD) simulations of the large scale jets \cite{McKinney12} considering magnetically-arrested disks (MAD) perturbing the jets. We also note that the QPOs observed in the source appear to be accompanied with lower frequency harmonics, similar to the low frequency `C-type' QPOs found in Galactic X-ray binaries. These QPOs are frequently interpreted in the context of the Lense-Thirring precession of the accretion disks \citep[see][]{Stella98,Motta11}. However, in our case the harmonics show lower significance over the red-noise processes.  Besides, as there are not many studies focused particularly on this  source, there are no robust estimates of its mass and the jet angle. Hence, it is hard to attribute the observed QPO conclusively to a singular scenario.

In the case of the detected QPO in the blazar PKS 0219 -164, the ratio of the total observation period to the observed primary period is (3053 d/270 d), more than 10. This implies that there are nearly 10 cycles of the quasi-periodic oscillations present in the source light curve. In addition, the significance estimated taking account of the fact the oscillations might be due to red-noise processes rather than the true periodicity is fairly large - global 99.6\% significance from the LSP and 97.5\% significance from the WWZ. The observed difference in the significances could be accounted for the fact that the WWZ method considers the transient nature of the QPO whereas the LSP method expects periodic oscillations to persist throughout the entire observation period. Moreover, during the course of the significance estimation in the WWZ method, the comparison is made between the source and the simulated variability powers which are averaged over the time; hence we lose some information. Therefore it might present one of the strongest cases of the detection of QPO in blazars. Finally, the analysis of the $\gamma$ ray observations were motivated to perform similar QPO analysis in the high energy, and thereby further investigate the location and nature of QPO in the blazar. However, due to smaller number of observations with TS$>$10, this could not be achieved.

\begin{acknowledgements}
I thank the anonymous referee for a careful review and important suggestions, which significantly improved the presentation of the paper. I acknowledge the financial support by the Polish National Science Centre through the grant DEC- 2012/04/A/ST9/00083.  I would like to express my gratitude to Prof. Staszek Zola for kindly allowing us to use his computational facility for this research work. I also  thank Prof. Micha{\l} Ostrowski  and  Prof. {\L}ukasz Stawarz for their useful comments and suggestions during the work. This research has made use of data from the OVRO 40-m monitoring program (Richards, J. L. et al. 2011, ApJS, 194, 29) which is supported in part by NASA grants NNX08AW31G, NNX11A043G, and NNX14AQ89G and NSF grants AST-0808050 and AST-1109911.

\end{acknowledgements}

\end{document}